\documentclass[conference]{IEEEtran}
\usepackage{amsmath,amssymb,amsfonts}
\usepackage{algorithmic}
\usepackage{multirow}
\usepackage{graphicx}
\usepackage{textcomp}
\usepackage{xcolor}
\usepackage{lipsum}
\usepackage{enumitem}
\usepackage{hyperref}
\usepackage[normalem]{ulem}

\makeatletter 
\newcommand{\linebreakand}{
  \end{@IEEEauthorhalign}
  \hfill\mbox{}\par
  \mbox{}\hfill\begin{@IEEEauthorhalign}
}

\usepackage[style=ieee]{biblatex}
\begin{document}

\title{Scalable and Consistent Graph Neural Networks for Distributed Mesh-based Data-driven Modeling}

\author{
\IEEEauthorblockN{Shivam Barwey\textsuperscript{\textsection}}
\IEEEauthorblockA{\textit{Transportation and Power Systems} \\
\textit{Argonne National Laboratory}\\
Lemont, USA \\
sbarwey@anl.gov}
\and
\IEEEauthorblockN{Riccardo Balin\textsuperscript{\textsection}}
\IEEEauthorblockA{\textit{Leadership Computing Facility} \\
\textit{Argonne National Laboratory}\\
Lemont, USA \\
rbalin@anl.gov}
\and
\IEEEauthorblockN{Bethany Lusch}
\IEEEauthorblockA{\textit{Leadership Computing Facility} \\
\textit{Argonne National Laboratory}\\
Lemont, USA \\
blusch@anl.gov}
\and
\linebreakand
\IEEEauthorblockN{Saumil Patel}
\IEEEauthorblockA{\textit{Computational Science Division} \\
\textit{Argonne National Laboratory}\\
Lemont, USA \\
spatel@anl.gov}
\and
\IEEEauthorblockN{Ramesh Balakrishnan}
\IEEEauthorblockA{\textit{Computational Science Division} \\
\textit{Argonne National Laboratory}\\
Lemont, USA \\
bramesh@anl.gov}
\and
\IEEEauthorblockN{Pinaki Pal}
\IEEEauthorblockA{\textit{Transportation and Power Systems} \\
\textit{Argonne National Laboratory}\\
Lemont, USA \\
pal@anl.gov}
\and
\linebreakand
\IEEEauthorblockN{Romit Maulik}
\IEEEauthorblockA{\textit{Information Sciences and Technology} \\
\textit{Pennsylvania State University}\\
University Park, USA \\
\textit{Mathematics and Computer Science} \\
\textit{Argonne National Laboratory}\\
Lemont, USA \\
rmaulik@psu.edu}
\and
\IEEEauthorblockN{Venkatram Vishwanath}
\IEEEauthorblockA{\textit{Leadership Computing Facility} \\
\textit{Argonne National Laboratory}\\
Lemont, USA \\
venkat@anl.gov}
}

\maketitle
\thispagestyle{plain}
\pagestyle{plain}
\begingroup\renewcommand\thefootnote{\textsection}
\footnotetext{Equal contribution.\\ \\ \textit{Accepted to the Proceedings of SC24-W: Workshops of the International Conference for High Performance Computing, Networking, Storage and Analysis.}}
\endgroup

\begin{abstract}
This work develops a distributed graph neural network (GNN) methodology for mesh-based modeling applications using a consistent neural message passing layer. As the name implies, the focus is on enabling scalable operations that satisfy physical consistency via halo nodes at sub-graph boundaries. Here, consistency refers to the fact that a GNN trained and evaluated on one rank (one large graph) is arithmetically equivalent to evaluations on multiple ranks (a partitioned graph). This concept is demonstrated by interfacing GNNs with NekRS, a GPU-capable exascale CFD solver developed at Argonne National Laboratory. It is shown how the NekRS mesh partitioning can be linked to the distributed GNN training and inference routines, resulting in a scalable mesh-based data-driven modeling workflow. We study the impact of consistency on the scalability of mesh-based GNNs, demonstrating efficient scaling in consistent GNNs for up to O(1B) graph nodes on the Frontier exascale supercomputer. 
\end{abstract}

\begin{IEEEkeywords}
graph neural networks, mesh-based modeling
\end{IEEEkeywords}

\section{Introduction}
\label{sec:intro}
The curation of large datasets of spatiotemporal physics, sourced from either high-fidelity canonical simulations or real-world/experimental sensor observations (e.g., advanced laser-based imaging), has in recent years opened up promising pathways for the development of data-driven models acting as surrogates or enhancers to conventional physics-based partial differential equation (PDE) simulation methods \cite{brunton_2024,pedram_closure_review}. Data-driven models for physics simulation applications are becoming increasingly more adopted due to the fact that their predictive capability necessarily scales with increases in size and quality of this data (such methods thrive in data-rich environments) \cite{mahoney_fm}. Ultimately, through the integration of these large datasets, data-driven models can be trained to eliminate prohibitive spatiotemporal restrictions arising in nonlinear PDEs, thereby opening up a direct pathway for vastly accelerated simulations \cite{romit_sciml}. In computational fluid dynamics (CFD), for example, the construction of orders-of-magnitude accelerated data-driven modeling frameworks is seeing significant spikes in popularity, leading to fast, GPU-compatible surrogates \cite{kochkov2021machine,romit_differentiable} that can be leveraged in various downstream engineering workflows. 

State-of-the-art data-driven approaches for accelerating physics simulations are built on neural network architectures inspired by those used in the machine learning and computer vision communities. Although many candidate neural network architectures have been (and continue to be) explored with good success, several approaches (such as convolutional neural networks \cite{gonzalez2018deep,fukami2020cnn} and transformer-type architectures \cite{zabaras_transformer,kang2023new}), though successful in modeling canonical physics, operate under the assumption of structured and fixed grid discretizations. This conflicts with the fact that practical data-driven models for physical simulations need to satisfy the critical \textit{complex geometry} requirement. High-fidelity simulations of practical systems are described by complex geometric features, such as intricacies in airfoil surfaces for aerodynamic applications and complex engine configurations in internal combustion / propulsion applications. As a result, raw high-quality training data for full-scale systems live on unstructured grids or meshes. Methods that are naturally compatible with these meshes are therefore required for data-driven models to become truly useful as a universal tool in industrial applications. 

The complex geometry requirement has led to increasing utilization of geometric deep learning \cite{gdl} concepts for \textit{mesh-based} simulation of physical systems, with graph neural networks (GNNs) as the modeling backbone. Since PDE solutions of complex geometries (e.g., velocity field in the case of the Navier-Stokes equations) are often defined on unstructured meshes, in mesh-based GNN modeling, the mesh is itself instantiated as the graph, allowing for GNN models to be executed and trained on high-fidelity data generated by multi-physics unstructured codes \cite{meshgraphnet,kolter_diffgnn}. In such GNNs, the graph nodes coincide with the spatial discretization/collocation points on which the solution is defined, and the edges connect neighboring nodes based on either pre-defined computational stencils or edge generation algorithms. The GNN modeling goal in the physical simulation context becomes one of node-level regression, e.g., predicting the velocity vectors on all nodes at some future time instant. 
{\color{black} Alongside node permutation-invariance}
{\color{black} (i.e., permutation of nodes with the same topology should
be considered the same graph),}
{\color{black} the key advantage garnered by this setup is that the same GNN model, once trained, can be applied to}
{\color{black} any} 
{\color{black} mesh-based graph, in the form of different meshes and geometries, during the inference stage \cite{chen_random}.}

Initial mesh-based GNNs found success using an encode-process-decode architecture \cite{meshgraphnet} based on \textcolor{black}{neural} message passing layers  \cite{gilmer}. This general strategy has evolved considerably in the past few years, particularly in the fluid dynamics community; improvements include incorporating multi-scale operations into \textcolor{black}{neural} message passing architectures \cite{multiscale_mgn,lino_2021,shivam_jcp,deshpande2024}, embedding knowledge of physics-based discretization rules into the graph connectivity \cite{karthik_gnn,jaiman_hypergraph}, modifying architectures for interpretability \cite{shivam_jcp}, and embedding physical equivariance/invariance properties into models to improve predictions \cite{lino_2021,varun_gnn}. Alongside purely unstructured grids, recent work has also extended GNN capabilities to produce models on moving and adaptive meshes \cite{gnn_amr,shivam_jcp}, which is another useful capability (supplementing complex geometry compatibility) provided by these architectures. 

Although previous studies have successfully demonstrated the advantages of mesh-based GNNs in terms of modeling physics on unstructured meshes, they are typically demonstrative in nature and leverage graph sizes on the order of $10^4$-$10^5$ nodes. Since the graph nodes here correspond to spatial discretization points, this translates to a relatively small problem size for \textcolor{black}{today's physics-based simulations.}
For example, in fluid dynamics simulations, developing more robust models requires interfacing with unstructured data generated by high-fidelity CFD solvers characterized by well-resolved unstructured meshes containing tens of millions (and higher) spatial discretization points \cite{prakash2024streamline,umrf,cascade_paper,nekrs}. This translates to individual graph sizes of O(10M) nodes and higher, with the large modern simulations executed on mesh sizes of 100M to 1B nodes. 
{\color{black} In addition to the inherent computing challenges stemming from the sheer size of these graphs}
{\color{black} (e.g., memory availability on current GPU),}
{\color{black} in PDE-based applications, such large graphs coincide with highly resolved meshes; the mesh-based data on these graphs therefore capture significantly more complex physical phenomena (e.g., turbulence) than what would appear on smaller under-resolved meshes, rendering an overall more challenging modeling task from the physical perspective.} 
Therefore, to realize the vision of robust mesh-based data-driven models, it is necessary to (a) develop new distributed \textcolor{black}{neural} message passing approaches that can scale to larger graph sizes, and (b) understand how critical parameters -- such as model size and graph size -- affect the scalability limits of GNN \textcolor{black}{neural} message passing operations during training and inference. 

To this end, this work develops a distributed GNN for mesh-based modeling applications that relies on novel alterations to the baseline \textcolor{black}{neural} message passing layer. These modifications facilitate scalable operations with emphasis on consistency. Here, consistency refers to the fact that a GNN trained and evaluated on one rank (on one large graph) is arithmetically equivalent to evaluations on multiple ranks (each of which operate on a sub-graph). {\color{black} This feature is particularly important for fluid dynamics and other PDE-based simulations where domain decomposition is used to tackle large problems and predicted solutions must be continuous across partition boundaries and unaffected by the partitioning scheme.}

The method introduced here is inspired by the way in which CFD solvers scale up their operations via domain decomposition. In any scalable physics simulation procedure, the mesh is partitioned into sub-domains, with alternating steps between arithmetically intense local state updates and bandwidth-limited non-local boundary communication (or synchronization) steps among the various ranks. Since the graph coincides with the mesh in mesh-based GNNs, the present approach leverages the \textit{same} domain decomposition information produced by the CFD code (i.e., the same partitioned mesh) in the GNN architecture, with alternating steps between local \textcolor{black}{neural} message passing on each sub-graph and non-local communication facilitated by halo exchanges among boundary graph node attributes. 

This concept is demonstrated here by interfacing graph neural networks (implemented in PyTorch \cite{pytorch} and PyTorch Geometric \cite{pygeom}) with NekRS \cite{nekrs}, a GPU-capable exascale CFD solver developed at Argonne National Laboratory. In particular, the mesh preprocessor in NekRS is used to provide the necessary sub-graph and halo exchange information required for consistent distributed GNN operations on the PyTorch side. To isolate the communication penalties introduced by these consistent \textcolor{black}{neural} message passing layers, an extensive scaling analysis is performed in various model and graph size configurations using the Frontier supercomputer at the Oak Ridge Leadership Computing Facility \cite{frontier}. 

It must be acknowledged that other recent works also explore distributed message passing GNNs in the context of mesh-based modeling \cite{kakka2024,scaling_gnn_scitech,nvidia_modulus}. We isolate the following contributions as distinctions from these works: (1) this work develops a consistent \textcolor{black}{neural} message passing layer that interprets the communication step as a modification to the baseline edge aggregation procedure; 
(2) we intentionally place less emphasis on the modeling task itself, and instead focus on the impact on scalability of enforcing consistency (and particularly differentiable halo exchanges) in GNN operations, demonstrating scaling up to $O(10^9)$ graph nodes on Frontier; and 
(3) we focus on interfacing with existing scalable simulation tools (i.e., NekRS and its associated domain-decomposed element-based meshing).

\section{Methods}
\label{sec:methods}
To solidify the modeling context for the methods developed here, the GNN operation is given by 
\begin{equation}
    \label{eq:gnn}
    {\bf Y}_r = {\cal G}({\bf X}_r, {\bf E}_r, {\bf A}_r; \theta), \quad r = 1,\ldots,R. 
\end{equation}
In Eq.~\ref{eq:gnn}, $\cal G$ represents the message passing-based graph neural network. The primary inputs are the node attribute matrix ${\bf X}_r \in \mathbb{R}^{N_x \times F_x}$, the edge attribute matrix ${\bf E}_r \in \mathbb{R}^{N_e \times F_e}$, and the graph adjacency matrix ${\bf A}_r \in \mathbb{Z}^{N_x \times N_x}$ which is stored in a sparse coordinate list format and contains the node connectivity information. The quantities $N_x$ and $F_x$ correspond to the number of graph nodes and node features respectively, while $N_e$ and $F_e$ correspond to the number of edges and edge features respectively. The output ${\bf Y}_r$ is a node attribute matrix -- as such, the model ${\cal G}$ predicts quantities at the node level. For example, when modeling fluid flows on a mesh, ${\bf X}_r$ may represent the velocity vectors at each node at some initial time $t_0$, with the GNN trained to predict velocities ${\bf Y}_r$ at a future time $t_f$ on the same mesh. Lastly, the quantity $\theta$ represents the GNN parameters.

The input and output values in Eq.~\ref{eq:gnn} are assigned the subscripts $r$, which designates a rank (or GPU) index out of a total of $R$ ranks participating in the GNN operations. The following two points are highlighted with respect to this rank-dependence: (1) when $R>1$, the adjacency matrices ${\bf A}_r$ (and associated node and edge attribute matrices) correspond to sub-graphs of a much larger graph; and (2) there is no dependency of the GNN \textit{parameters} $\theta$ on the rank index, i.e., all ranks leverage the same model $\cal G$ to generate predictions during training and inference (in the case of training, this resembles a distributed data parallel setup). 

Each rank therefore produces the output ${\bf Y}_r$ on its own subgraph, which is interpreted as a subset of a much larger mesh on which the graph nodes are defined. This interpretation, detailed in Sec.~\ref{sec:graphgen}, requires constructing a graph from a domain decomposed mesh, and is equivalent to a spatially-parallel approach for domain decomposition. Additionally, although the parameters $\theta$ are shared across all ranks, to facilitate consistent GNN operations during training and inference, the specific evaluations internal to the GNN $\cal G$ (the \textcolor{black}{neural} message passing layers) are conditioned on the present rank index $r$ as well as neighboring ranks to enforce consistency in messages propagated across sub-graph boundaries. The consistent \textcolor{black}{neural} message passing layer, which leverages halo nodes for this purpose, is detailed in Sec.~\ref{sec:consistent_mp}.

\begin{figure}
    \centering
    \includegraphics[width=0.8\linewidth]{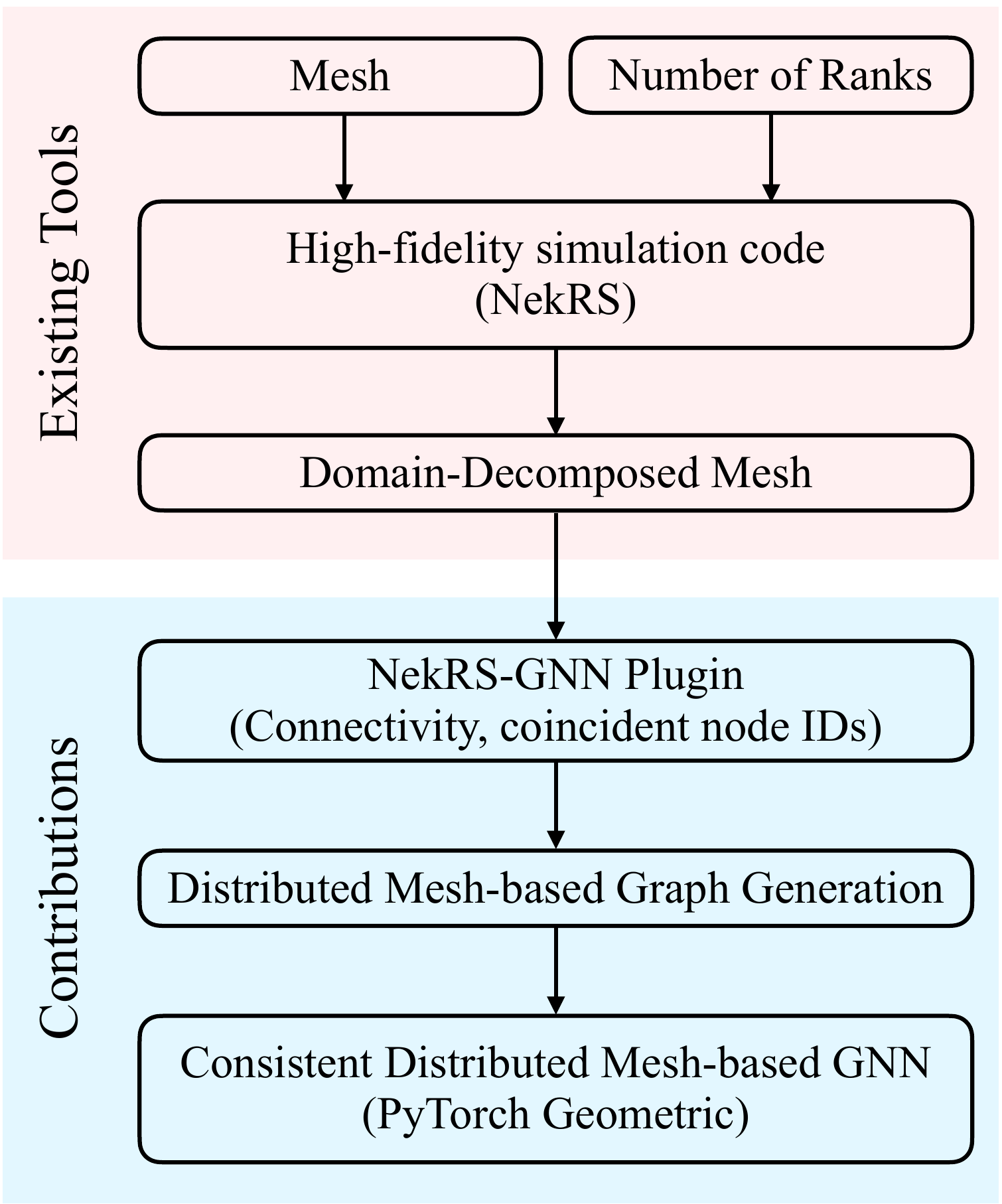}
    \caption{Overview of workflow, with components from existing tools highlighted in red and contributions of this work highlighted in blue. Code for the NekRS interface and consistent GNN implementation is openly available {\color{black} at the following GitHub repository: \url{https://github.com/argonne-lcf/nekRS-ML/tree/GNN}}.} 
    \label{fig:overview}
\end{figure}

Figure~\ref{fig:overview} provides an overview of the various steps involved in the methodology to be described in the subsections below. Given a mesh and number of decomposition ranks as a starting point (top boxes in Fig.~\ref{fig:overview}), the CFD solver provides the domain-decomposed mesh. This decomposed mesh is then parsed by a \verb|NekRS-GNN| plugin that interfaces with the solver mesh object to extract graph connectivity information and node ID information required to generate the mesh-based graph. Upon generation of the distributed mesh-based graph (which leverages halo nodes), a distributed GNN is then constructed through the utilization of the consistent \textcolor{black}{neural} message passing layers. From the GNN implementation side, emphasis is placed here on compatibility with PyTorch communication libraries and PyTorch Geometric formats -- as such, "graph generation" refers to the construction of a distributed graph representation in PyTorch Geometric from the corresponding mesh utilized in the CFD solver.

\subsection{Distributed Mesh-based Graph Generation}
\label{sec:graphgen}
Graphs in this work coincide with element-based mesh discretizations of the so-called ``physical space" (a three-dimensional Cartesian space). Such element-based discretizations are used, as the names imply, in finite element, spectral element, and finite volume numerical methods for solving PDEs. It should be noted that, although the discussion below is geared towards generating graphs produced by NekRS discretizations, the distributed GNN concepts covered here apply to any mesh composed by a collection of finite elements and consumed by any code or tool solving a set of discretized governing PDEs.

\begin{figure}
    \centering
    \includegraphics[width=\linewidth]{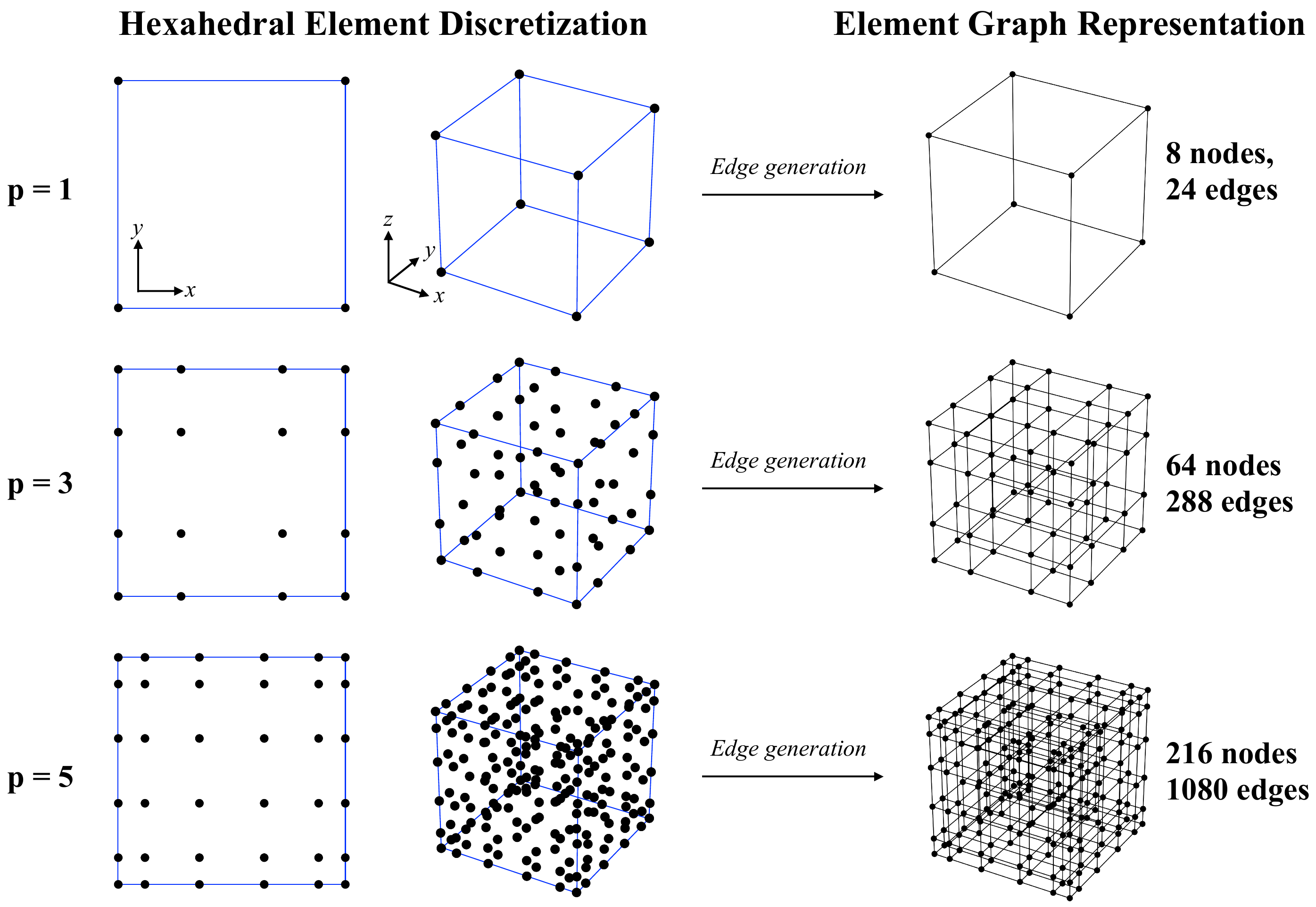}
    \caption{Illustration of element-based discretizations and graph generation. Left plots show elements at increasing polynomial orders per the Gauss-Legendre-Lobato (GLL) quadrature of NekRS \cite{nekrs}, with black markers denoting spatial quadrature points and blue lines denoting element boundaries. Right plots show corresponding graph representations produced after taking quadrature points as nodes (black markers) and generating edges (black lines) to connect neighboring nodes.}
    \label{fig:element_graphgen}
\end{figure}

NekRS leverages the spectral element method, and therefore discretizes the spatial domain with a mesh composed of non-intersecting elements (overlapping overset meshes are not considered for this work). Each element in the domain is characterized by a \textit{polynomial order}, which prescribes the number of spatial quadrature points within a single element at which the quantities of interest (e.g., time-evolving velocity, pressure and temperature) are computed, as shown in Fig.~\ref{fig:element_graphgen} (left) for two and three dimensional elements. The arrangement of quadrature points in physical space within each element follows the Gauss-Legendre-Lobato (GLL) quadrature rule, with higher polynomial orders providing more refined discretization within each element and generally more accurate PDE solutions. Although the details of the solution accuracy with increasing polynomial order are out-of-scope here, the main takeaways are that (a) the GLL rule produces a non-uniform arrangement of the discretization points in physical space within each element, (b) the number of quadrature points per element scales according to $(p+1)^3$ for a given polynomial order $p$, and (c) the representation of the modeled QoIs in finite/spectral element space is dependent on both the shape of the mesh elements (in general, NekRS can handle mixed unstructured mesh elements consisting of wedges, tetrahedra, and hexahedra) as well as the polynomial order, resulting in a highly unstructured discretization. 

In the mesh-based GNN modeling approach, spatial discretization points coincide with the nodes of the graph. As such, in the graph generation strategy leveraged here, the spatial quadrature points defined within each element are instantiated as the graph nodes. Given this point-cloud collection of graph nodes, each of which is characterized by a three-dimensional spatial position attribute (x, y, and z coordinates), there are many ways to generate graph connectivities or edges. Here, undirected edges are created to connect neighboring quadrature nodes. The result of this approach is shown in Fig.~\ref{fig:element_graphgen} (right), where three increasing polynomial orders ($p=1,3,5$) are depicted to highlight how the graphs for these elements change as the level of refinement within each element increases. In the $p=1$ case, the graph edges are equivalent to the mesh edges, with graph nodes coinciding with hexahedral element vertices. At polynomial orders of $p>1$, the addition of more graph nodes inside the element produces a more refined graph connectivity with smaller average edge lengths. 

\begin{figure*}
    \centering
    \includegraphics[width=\textwidth]{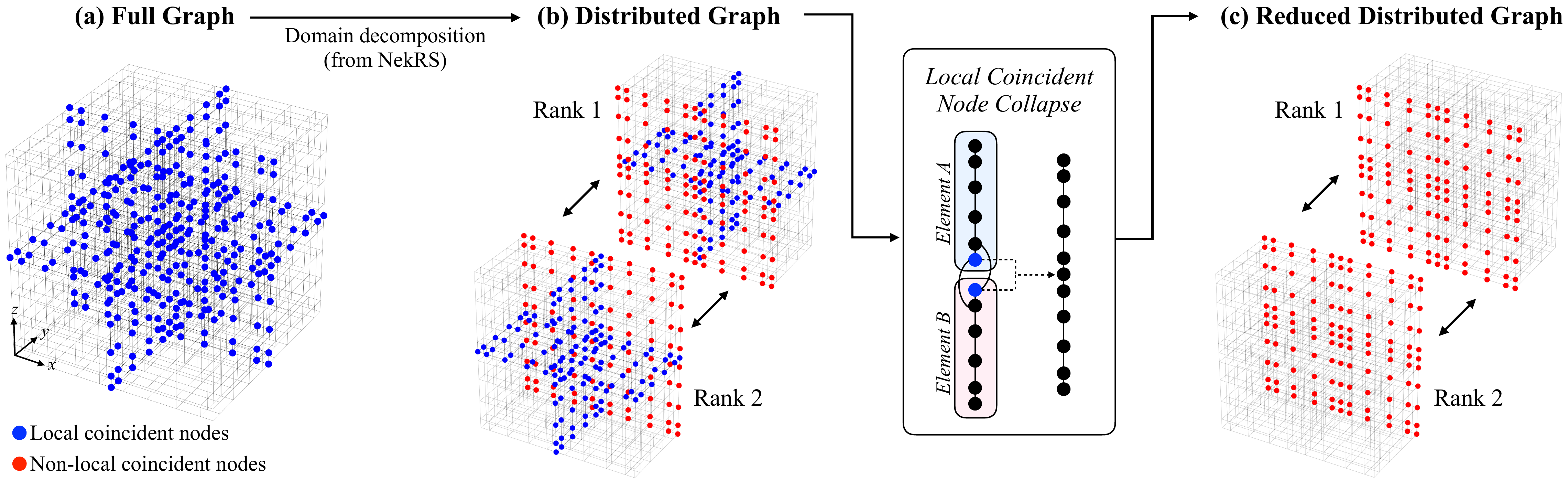}
    \caption{\textbf{(a)} Full $R=1$ graph of mesh composed of 8 total elements, each characterized by polynomial order $p=5$ (refer to Fig.~\ref{fig:element_graphgen}). Blue markers indicate local coincident nodes and element boundaries -- all other nodes are not shown for ease of visualization. \textbf{(b)} Corresponding distributed $R=2$ graph, highlighting the construction of non-local coincident nodes (red markers). Arrows between sub-graphs indicate communication directions required to enforce consistency at these nodes. \textbf{(c)} Reduced distributed graph produced by collapsing/consolidating local coincident nodes. Block between (b) and (c) illustrates the node collapse procedure between two neighboring elements on a local graph in 1D.}
    \label{fig:dist_graphgen}
\end{figure*}

Using the element-based graph representation described above as the building block, distributed graphs of full meshes, which are composed of many such non-intersecting elements (interpreted as a concatenation of many smaller element-based graphs), can be generated by leveraging the domain decomposition routines of NekRS. The full procedure is shown in Fig.~\ref{fig:dist_graphgen}. Figure~\ref{fig:dist_graphgen}(a) shows a simplified one-rank mesh ($R=1$) consisting of 2 elements in each spatial direction (8 elements in total). The figure highlights the presence of \textit{local coincident nodes}, which is an artifact of element-based meshes: nodes on the faces of neighboring elements are coincident, and therefore share the same physical position. In the graph representation, node neighbors among all coincident nodes are shared. The blue markers in Fig.~\ref{fig:dist_graphgen}(a) highlight these local coincident nodes, which also serve as markers for the boundaries between the 8 elements. Coincident nodes are formally extracted using global node indices, with two or more nodes sharing the same global node index defined as coincident. 

The NekRS-provided decomposition of the "full" mesh-based graph shown in Fig.~\ref{fig:dist_graphgen}(a) into a distributed graph stored on two ranks ($R=2$) is shown in Fig.~\ref{fig:dist_graphgen}(b). The figure highlights how distribution of graph elements onto multiple ranks results in the formation of \textit{non-local} coincident nodes, indicated by the red markers. The presence of a non-local coincident node on Rank 1, for example, signifies the existence of at least one other graph node with the same global index (same spatial coordinate) on another rank (in this case, Rank 2). Shown in Fig.~\ref{fig:dist_graphgen}(c) is a reduced distributed graph representation that is used for GNN operations. A reversible indexing representing the consolidation (or collapse) of local coincident nodes to one owner (shown in the box in Fig.~\ref{fig:dist_graphgen}) can be used to eliminate duplicate nodes on local graphs, reducing the overall local node count on each rank. Since this eliminates the need to execute local synchronization steps, the final distributed graph representation shown in Fig.~\ref{fig:dist_graphgen}(c) -- referred to as the "reduced" distributed graph -- contains only non-local coincident nodes. 

Physical consistency requires node attributes on all coincident nodes (both local and non-local) to be identical. Enforcement of this constraint on the local coincident nodes is handled implicitly by means of the reduced distributed graph. On the other hand, enforcement of this constraint on non-local coincident nodes requires communication between ranks, and motivates the construction of the consistent \textcolor{black}{neural} message passing layer (to be described in Sec.~\ref{sec:consistent_mp}) facilitated by halo nodes and associated halo exchanges between ranks.

\begin{figure}
    \centering
    \includegraphics[width=0.8\linewidth]{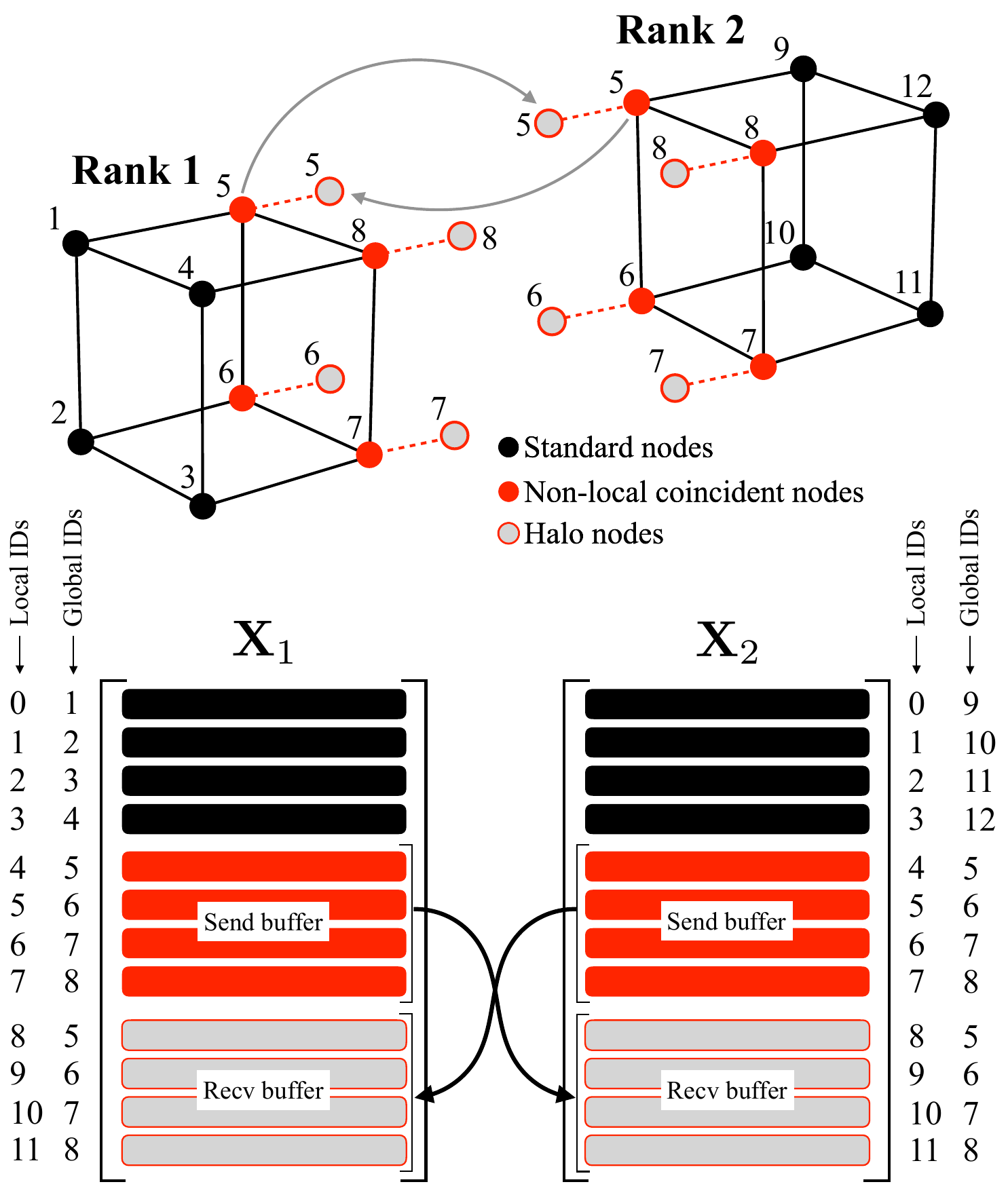}
    \caption{\textbf{(Top)} Schematic of halo nodes for an $R=2$ distributed graph consisting of two $p=1$ elements. \textbf{(Bottom)} Visualization of node attribute matrices involved in the exchange.}
    \label{fig:halo_schematic}
\end{figure}

An illustration of the halo nodes and exchange procedure is shown in Fig.~\ref{fig:halo_schematic} for a simplified configuration consisting of a distributed graph of two $p=1$ elements partitioned into two ranks ($R=2$, one element per rank). The halo nodes (gray markers) receive attribute values from non-local coincident nodes at the same global ID in the neighboring rank. Depicted in Fig.~\ref{fig:halo_schematic} is also an illustration of node attribute matrices involved in the exchange. The additional halo nodes are appended to the node attribute matrix, with send and receive buffers corresponding to coincident (red) and halo (gray) subsets of rows, respectively. To execute the exchange on a given rank, index send masks are used to first extract the send buffers (subsets of the note attribute matrix) for the respective neighboring ranks. Receive masks are then used to update the halo node attributes on the receiving rank. The size of the buffer involved in the exchange is equivalent to the number of halo nodes multiplied by the node feature dimensionality. In the \textcolor{black}{neural} message passing layer described below, the halo nodes are used to facilitate exchanges over aggregated edge features (i.e., ${\bf X}_1$ and ${\bf X}_2$ in Fig.~\ref{fig:halo_schematic} represent local neighborhood aggregates) to ensure that messages in local graphs consistently propagate to neighboring ranks. This occurs under a communication cost penalty proportional to the buffer size, which is the focus of the scalability analysis in Sec.~\ref{sec:results}.  

\subsection{Consistent Neural Message Passing Layer}
\label{sec:consistent_mp}
This section provides a description of the consistent \textcolor{black}{neural} message passing (NMP) layer. Consistency in this context means that GNN (and NMP layer) outputs should be (a) invariant to the number of ranks, and (b) equivalent to what would be produced by the baseline $R=1$, or un-partitioned graph. 

More formally, the consistency requirement can be described by
\begin{equation}
    \label{eq:consistency_1}
    {\cal S}\left(\{ {\bf Y}_1^{\text{local}} \}_{R=1} \right) = {\cal S}\left( \text{cat}\{ {\bf Y}_1^{\text{local}}, \ldots,  {\bf Y}_R^{\text{local}} \}_{R>1} \right)
\end{equation}
and
\begin{equation}
    \label{eq:consistency_2}
    \frac{\partial {\cal S}}{\partial \theta} \bigg\rvert_{R=1} = \frac{\partial {\cal S}}{\partial \theta} \bigg\rvert_{R>1}. 
\end{equation}

In Eq.~\ref{eq:consistency_1}, $\cal S$ represents an arbitrary function (either linear or nonlinear, scalar or vector-valued) operating on the graph nodes. The left-hand-side corresponds to the function values produced by the $R=1$ (un-partitioned) graph, and the right-hand side corresponds to the values produced by the $R>1$ partitioned graph. The inputs to the functions are the GNN output node atribute matrices ${\bf Y}_r^{\text{local}}$, with the superscript ``$\text{local}$" indicating that rows corresponding to halo nodes are discarded. The ``$\text{cat}$" operation in Eq.~\ref{eq:consistency_1} is a row-wise concatenation of the set of node attribute matrices in the $R>1$ case (with all nodes assumed to be ordered according to their global indices). Ultimately, Eq.~\ref{eq:consistency_1} states that any function evaluated on the GNN output must be independent of the graph partitioning strategy, which is a physically consistent requirement in the context of mesh-based modeling. 

As an example, $\cal S$ can be interpreted as a scalar-valued loss function used in training (this so-called consistent loss function is described in Sec.~\ref{sec:consistent_loss}). As shown in Eq.~\ref{eq:consistency_2}, a fully consistent formulation also requires gradients of the function $\cal S$ with respect to some GNN parameter $\theta$ to be invariant to the graph partitioning scheme. In other words, GNN gradients produced for partitioned graphs in the $R>1$ case should align with those computed in the un-partitioned $R=1$ case. The implication of enforcing such consistency is that training convergence rates and model sensitivities are invariant to both the number of partitions and the partition boundaries themselves for a given problem. Enforcing Eq.~\ref{eq:consistency_1} requires incorporating halo exchanges in the \textcolor{black}{neural} message passing layer; enforcing Eq.~\ref{eq:consistency_2} in the context of distributed training requires utilizing \textit{differentiable} communication routines. 

The consistent NMP layer builds upon a "standard" NMP layer format that has been successful in previous mesh-based data-driven modeling works \cite{meshgraphnet,lino_2021,shivam_jcp}. The standard NMP layer involves a three-stage process: (1) an edge feature update conditioned on its corresponding sender and receiver nodes, (2) a summation-aggregation of the updated edge features, and (3) a node feature update conditioned on the aggregated edge features. The consistent NMP layer adds an additional set of steps before the final node update in the form of halo swaps and synchronization of the edge aggregates. The equations for the consistent NMP layer on a given rank $r$ sub-graph are as follows: 
\begin{subequations}
\begin{align}
\text{\underline{Edge update}}&: {\bf e}_r^{ij} = \text{MLP}({\bf x}_r^{i}, {\bf x}_r^{j}, {\bf e}_r^{ij}), \label{eq:cmp_edge_update}\\
\text{\underline{Local edge aggr.}}&: {\bf a}_r^{i} = \sum_{j \in N(i)} \frac{1}{d_r^{ij}} {\bf e}_r^{ij}, \label{eq:cmp_edge_aggr}\\
\text{\underline{Halo swap}}&: {\bf a}_r^{i,\text{halo}} = {\bf a}_s^{k,\text{local}}  \text{ if } G_r(i) = G_s(k), \label{eq:cmp_halo_swap}\\
\text{\underline{Synchronization}}&: {\bf a}_r^{i,*} = \sum_{\substack{j\\G_r(j)=G_r(i)}} {\bf a}_r^j, \label{eq:cmp_sync}\\
\text{\underline{Node update}}&: {\bf x}_r^{i} = \text{MLP}({\bf a}_r^{i,*}, {\bf x}_r^{i}). \label{eq:cmp_node_update}
\end{align}
\label{eq:cmp_layer}
\end{subequations}

The consistent NMP layer begins in conventional fashion with Eq.~\ref{eq:cmp_edge_update}, which is the edge feature update step. Here, the edge features for an edge connecting nodes $i$ and $j$ on the rank $r$ sub-graph are given by ${\bf e}^{ij}_r$. The edge update leverages a multi-layer perceptron (MLP) with residual connections, and takes as input the corresponding node features ${\bf x}_r^i$ and ${\bf x}_r^j$ alongside the previous input edge features. 

The updated edge features corresponding to receiver node $i$ are then \textit{locally} aggregated in Eq.~\ref{eq:cmp_edge_aggr}. This aggregation comes from a summation of edge features in the neighborhood of node $i$, with the set of neighboring node indices given by $N(i)$. During aggregation, edge features are scaled by the inverse of the edge degree $d_r^{ij}$ to account for the presence of duplicate edges at the boundary interface of sub-graphs. $d_r^{ij} > 1$ if both nodes $i$ and $j$ are non-local coincident nodes (e.g., in the Fig.~\ref{fig:halo_schematic} schematic, the edges on the shared face connecting the red nodes have degrees of 2 -- scaling by this degree allows the final aggregated values to be consistent with what would be computed on the un-partitioned $R=1$ graph).

After the local edge aggregation in Eq.~\ref{eq:cmp_edge_aggr}, a halo swap (Eq.~\ref{eq:cmp_halo_swap}) is executed to populate halo nodes with aggregated edge features from neighboring ranks. In Eq.~\ref{eq:cmp_halo_swap}, ${\bf a}_r^{i,\text{halo}}$ is the aggregate at the halo node index $i$ for owner rank $r$, and ${\bf a}_s^{k,\text{local}}$ is the local aggregate at node index $k$ for a graph on a neighboring rank $s$. The exchange only proceeds for nodes that share the same global identifiers (i.e., non-local coincident nodes); $G_r(i)$ is the global node index for the $i$-th node on the rank $r$ sub-graph, and $G_s(k)$ is the same quantity for the $k$-th node on rank $s$. As illustrated in Fig.~\ref{fig:halo_schematic}, the halo exchange is executed in the standard fashion using masks to populate send/receive buffers as subsets of node attribute matrices on the participating ranks. The synchronization step in Eq.~\ref{eq:cmp_sync} then sets the final aggregated quantity ${\bf a}_r^{i,*}$ as the summation of all aggregates on nodes (including halo nodes) sharing the same global index. For unique (non-coincident) nodes, Eqs.~\ref{eq:cmp_halo_swap} and Eqs.~\ref{eq:cmp_sync} are not utilized, and ${\bf a}_r^{i,*} = {\bf a}_r^{i}$. For coincident nodes with halo nodes (e.g., red nodes in Fig.~\ref{fig:dist_graphgen}), ${\bf a}_r^{i,*} \neq {\bf a}_r^{i}$, and Eqs.~\ref{eq:cmp_layer}b-d constitutes a \textit{non-local} and consistent edge aggregation. 

\begin{figure*}
    \centering
    \includegraphics[width=\textwidth]{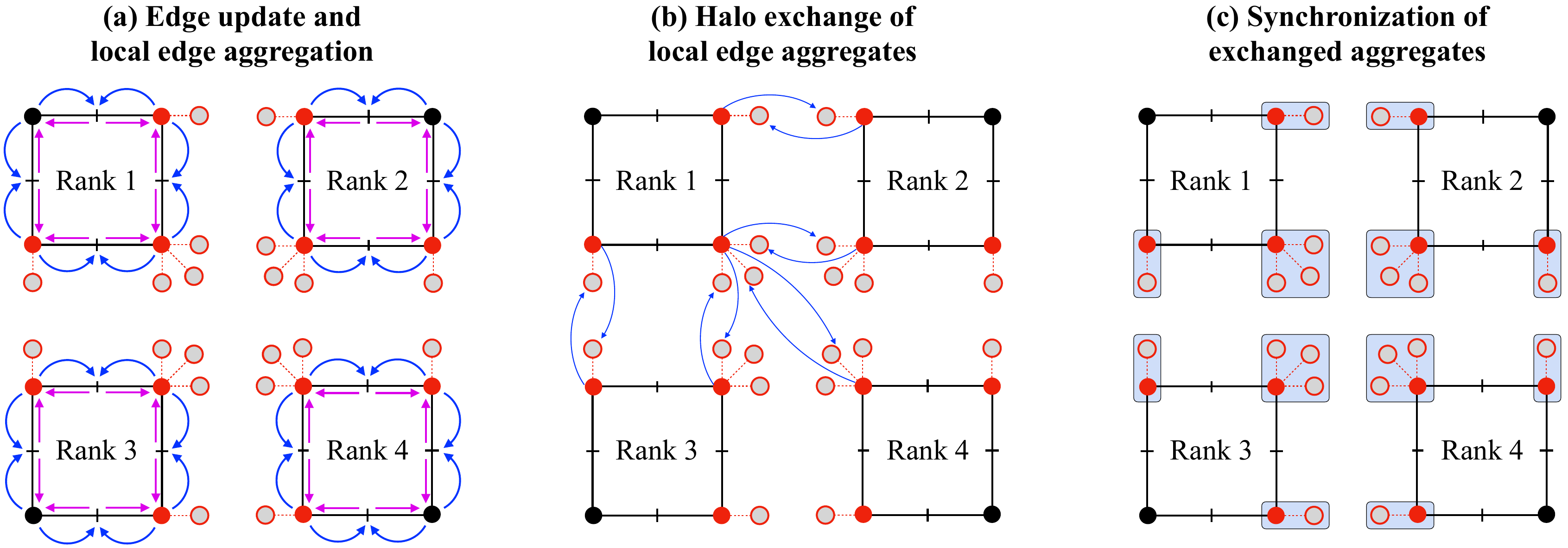}
    \caption{Visualization of consistent NMP layer steps on a small $R=4$ distributed graph composed of four $p=1$ elements for illustrative purposes. 2D projections are shown for ease of visualization. Node colors are same as Fig.~\ref{fig:halo_schematic} legend. \textbf{(a)} Illustration of edge update (blue arrows, Eq.~\ref{eq:cmp_edge_update}) and local edge aggregation (purple arrows, Eq.~\ref{eq:cmp_edge_aggr})  \textbf{(b)} Halo exchange over the local edge aggregates between neighboring ranks, populating the respective halo nodes (Eq.~\ref{eq:cmp_halo_swap}). Swap directions shown in blue arrows for exchanges involving Rank 1 only for visual clarity. \textbf{(c)} Synchronization step of the exchanged aggregates (Eq.~\ref{eq:cmp_sync}). Summations of the aggregated features ${\bf a}_r^i$ occur locally among coincident-halo node pairs sharing the same global index (indicated by regions in blue boxes).}
    \label{fig:cmp_layer}
\end{figure*}

The layer concludes with a node update MLP in Eq.~\ref{eq:cmp_node_update}, which updates node features using the aggregates ${\bf a}_r^{i,*}$ and input features ${\bf x}_r^i$. An illustration of the consistent NMP layer is provided in Fig.~\ref{fig:cmp_layer}. It can be shown that the formulation in Eq.~\ref{eq:cmp_layer} satisfies consistency in the output with respect to Eq.~\ref{eq:consistency_1}. So long as the halo exchange routine used in Eq.~\ref{eq:cmp_halo_swap} is differentiable, the formulation also satisfies consistency with respect to Eq.~\ref{eq:consistency_2} during backpropagation (discussed in Sec.~\ref{sec:results}). 

{\color{black} Lastly, it is emphasized that although the consistent NMP layer is interpreted as a modification to the standard NMP aggregation step, the integration of halo nodes as described here can be generally applied to extend non-local operations in other layers (e.g., attention layers over nodes}
{\color{black} or convolutions)}
{\color{black} to satisfy the consistency property.}

\subsection{Consistent Loss Function}
\label{sec:consistent_loss}

Training strategies in mesh-based data-driven models for physics simulations typically leverage mean-squared error (MSE) losses on the output node attribute matrices with respect to a target. On the un-partitioned graph ($R=1$), this becomes
\begin{equation}
    \label{eq:loss_standard}
    {\cal L} = \text{MSE}({\bf Y}_1, \widehat{\bf Y}_1) = \frac{1}{N F_y} \sum_{i=1}^{N} \sum_{j=1}^{F_y} ( {\bf Y}_1^{(i,j)} - \widehat{\bf Y}_1^{(i,j)} )^2,
\end{equation}
where $N$ is the number of nodes, $F_y$ the number of output features per node, and $\widehat{\bf Y}_1$ the target node attribute matrix. 

{\color{black} In the case of standard distributed data parallel training with $R>1$, independently evaluating local MSEs on each sub-graph using Eq.~\ref{eq:loss_standard} leads to an inconsistent formulation of the loss (i.e., aggregated local MSEs are not equivalent to the $R=1$ counterpart, violating Eq.~\ref{eq:consistency_1}). To mitigate this, a consistent loss in the $R>1$ scenario can be computed as follows: }

\begin{subequations}
\begin{align}
{\cal L} &= \frac{1}{N_{\text{eff}} F_y} \text{AllReduce}({\cal S}_r), \quad r = 1,\ldots,R, \text{ where} \label{eq:loss_consistent_a}\\
{\cal S}_r &= \sum_{i=1}^{N_r^{\text{local}}} \sum_{j=1}^{F_y} \frac{1}{d^i_r} ({\bf Y}_r^{(i,j)} - \widehat{\bf Y}_r^{(i,j)})^2, \text{ and} \label{eq:loss_consistent_b}\\
N_{\text{eff}} &= \text{AllReduce}(N_r), \quad N_r = \sum_{i=1}^{N_r^{\text{local}}} \frac{1}{d_r^i}. \label{eq:loss_consistent_c}
\end{align}
\label{eq:loss_consistent}
\end{subequations}

Equation~\ref{eq:loss_consistent_a} is the consistent MSE loss, and recovers Eq.~\ref{eq:loss_standard} when $R>1$. There are two AllReduce calls involved. The first is on the sum of squared-errors on the local graphs, denoted ${\cal S}_r$, which is defined in Eq.~\ref{eq:loss_consistent_b}. The sum is taken over the number of local nodes $N_r^{\text{local}}$ on each sub-graph (halo nodes are discarded). Squared errors are scaled by the inverse node degree $d_r^i$ to avoid double-counting the contribution of coincident nodes on different ranks. The second AllReduce recovers the ``effective" number of graph nodes $N_\text{eff}$, which is shown in Eq.~\ref{eq:loss_consistent_c}. The reduction occurs over a summation of the inverse node degrees, which ensures all ranks see an $N_{\text{eff}}$ that is equivalent to the un-partitioned number of nodes $N$ used in Eq.~\ref{eq:loss_standard}. 
\textcolor{black}{It is worth noting that, while Eq.~\ref{eq:loss_standard} is the standard MSE loss available in many ML framework libraries, such as PyTorch, Eq.~\ref{eq:loss_consistent} must be implemented as a customized loss function with access to a communication library, such as PyTorch Distributed.}

\section{Results}
\label{sec:results}
With the methods described above, the goal of this section is to (a) demonstrate and verify the consistency aspects of a GNN architecture composed of consistent \textcolor{black}{neural} message passing layers (Sec.~\ref{sec:consistency_demo}), and (b) provide a detailed scaling analysis to isolate the penalties incurred by a consistent model versus an inconsistent model (with and without halo exchanges respectively, Sec.~\ref{sec:scaling}). 
\textcolor{black}{The code for the NekRS interface and consistent GNN implementation used to obtain the following results is openly available at the GitHub repository \url{https://github.com/argonne-lcf/nekRS-ML/tree/GNN}}.

Prior to conducting this analysis, a GNN architecture for Eq.~\ref{eq:gnn} must be specified. The architectures used here follow the vetted encode-process-decode approach used in many previous mesh-based GNN modeling works \cite{meshgraphnet,lino_2021,shivam_jcp}. Briefly, the architecture is composed of the following components:
\begin{enumerate}[leftmargin=*]
    \item \textbf{Node and edge encoder: }The input node and edge features are lifted from the input feature size ($N_x$ and $N_e$ for nodes and edges respectively) to a hidden channel dimension $N_H$ using separate MLPs. These MLPs are purely local, meaning there are no rank-to-rank communication steps in the encoder phase. 
    \item \textbf{Neural Message passing: }A set of $M$ \textcolor{black}{neural} message passing layers are then used to model information exchange in the graph node neighborhoods. Here, the conventional NMP layers are replaced with the consistent formulation described in Sec.~\ref{sec:consistent_mp}. 
    \item \textbf{Node decoder: }After message passing, the hidden node features are transformed to an output feature dimensionality using another local MLP (a reversal of the encoding stage). Edge features are discarded, since loss functions (see Sec.~\ref{sec:consistent_loss}) are typically specified at the node level in mesh-based GNN modeling workflows. 
\end{enumerate}

Two GNN configurations -- referred to as ``small" and ``large" -- are constructed to understand the effect of critical parameters on the scaling performance of the consistent \textcolor{black}{neural} message passing operation. The details of both configurations are provided in Table~\ref{table:models}. MLPs used throughout GNNs leverage residual connections with layer normalization and ELU activation functions. Input node feature dimensionality is 3 (with node features representing three components of fluid velocity), and input edge dimensionality is 7, with edge features initialized using relative node features (three components), node distance vectors (three components), and magnitudes of distance vectors (one component).

\begin{table}[!ht]
  \centering
  \caption{Small and large GNN model settings.}
  \begin{tabular}{|p{4.2cm}||p{1.5cm}|p{1.5cm}|}
    \hline
    \textbf{GNN Description}& \textbf{Small} & \textbf{Large} \\ \hline
    \hline
    \textbf{Hidden channel dim. ($N_H$)} & 8 & 32 \\ \hline
    \textbf{Neural message passing layers ($M$)} & 4 & 4 \\ \hline
    \textbf{MLP hidden layers} & 2 & 5\\ \hline
    \textbf{Trainable parameters} & 3,979 & 91,459 \\ \hline
    \textbf{Halo exchange modes} & None, A2A, N-A2A & None, A2A, N-A2A \\ \hline
    \textbf{Nodes-per-subgraph/GPU} & 256k, 512k & 256k, 512k \\ \hline
  \end{tabular}
  \label{table:models}
\end{table}

The GNN architectures are implemented in PyTorch and PyTorch Geometric. Distributed training across multiple ranks/GPUs is performed with PyTorch Distributed Data Parallel (DDP). Central to the consistent NMP layer and scalability of the distributed GNN is the halo exchange implementation; in this work, halo exchanges are implemented using routines in the \verb|torch.distributed.nn| library. The communication routines in this library add differentiable functionality (i.e., gradient backpropagation capability) to a subset of baseline functions available in the \verb|torch.distributed.c10d| library, which is required for consistency in the training phase (refer to Eq.~\ref{eq:consistency_2}). Four halo exchange ``implementations" are compared:
\begin{itemize}[leftmargin=*]
    \item \textbf{No exchange (None):} This mode ignores the halo exchange step (Eqs.~\ref{eq:cmp_halo_swap} and \ref{eq:cmp_sync} are ignored in the consistent NMP layer), leading to an \textit{inconsistent} formulation using a conventional NMP layer. Since there is no halo exchange communication involved in this setting, this provides a baseline by which communication overhead deriving from consistency in the forward and backward passes can be assessed. Note, however, that the consistent loss in Eq.~\ref{eq:loss_consistent} is still computed, therefore three (two in the forward and one in the backward passes) additional AllReduce operations are present on top of the standard reduction on the gradients. 
    \item \textbf{AllToAll (A2A):} This mode utilizes the \verb|all_to_all| function in the \verb|torch.distributed.nn| library with standard buffer configurations (i.e., equal-sized buffers). Here, all $R$ ranks communicate with each other in fully connected fashion with the same buffer size, regardless of whether or not communication is needed. Although this is expected to be sub-optimal (since ranks that do not share halo nodes are communicating ``dummy" buffers), this provides a baseline differentiable halo exchange approach with which to assess more optimized configurations. 
    \item \textbf{Neighbor-AllToAll (N-A2A):} This mode utilizes the same \verb|all_to_all| function as above, but initializes the buffers in a selective manner to exploit a lesser-known functionality of  \verb|torch.distributed.nn| and vendor collective libraries (i.e., NCCL, RCCL or oneCCL for NVIDIA, AMD, and Intel GPU, respectively). More specifically, the lists of send and receive buffer tensors are filled with pre-allocated PyTorch Tensors only for the ranks that indeed share halo nodes, as identified by the NekRS mesh partitioner. For the remaining ranks not involved in the halo exchange, empty Tensors created with \verb|torch.empty(0)| are added to the buffer lists. The use of empty tensors informs the collective communication library set as the backend to PyTorch DDP to skip the communication between ranks that do not share halo nodes, hence reducing the \verb|all_to_all| to communication pattern similar to a \verb|send_receive| between neighbors. It should be noted that to the best of the authors' knowledge, this setting is not explicitly documented in the \verb|all_to_all| function description in PyTorch. \textcolor{black}{Lastly,} note that ``neighbor” in this context refers to neighboring ranks (ranks that share sub-graph boundaries and halo nodes), not to be confused with neighbors of some node in a local graph.  
    \item \textbf{Custom Send-Receive (Send-Recv):} The authors also experimented with a custom differentiable halo exchange approach using the \verb|isend| and \verb|irecv| primitives available in the \verb|torch.distributed.c10d| library. This approach is similar to N-A2A, in the sense that only neighboring ranks communicate with each other, with non-neighboring rank communication ignored. Due to space limitations, results using this custom implementation are not reported here. In summary, we found this implementation to be less robust to large buffer sizes and updates to the versions of PyTorch and the collective libraries. 
\end{itemize}

\subsection{Demonstration of Consistency}
\label{sec:consistency_demo}

\begin{figure}
    \centering
    \includegraphics[width=\linewidth]{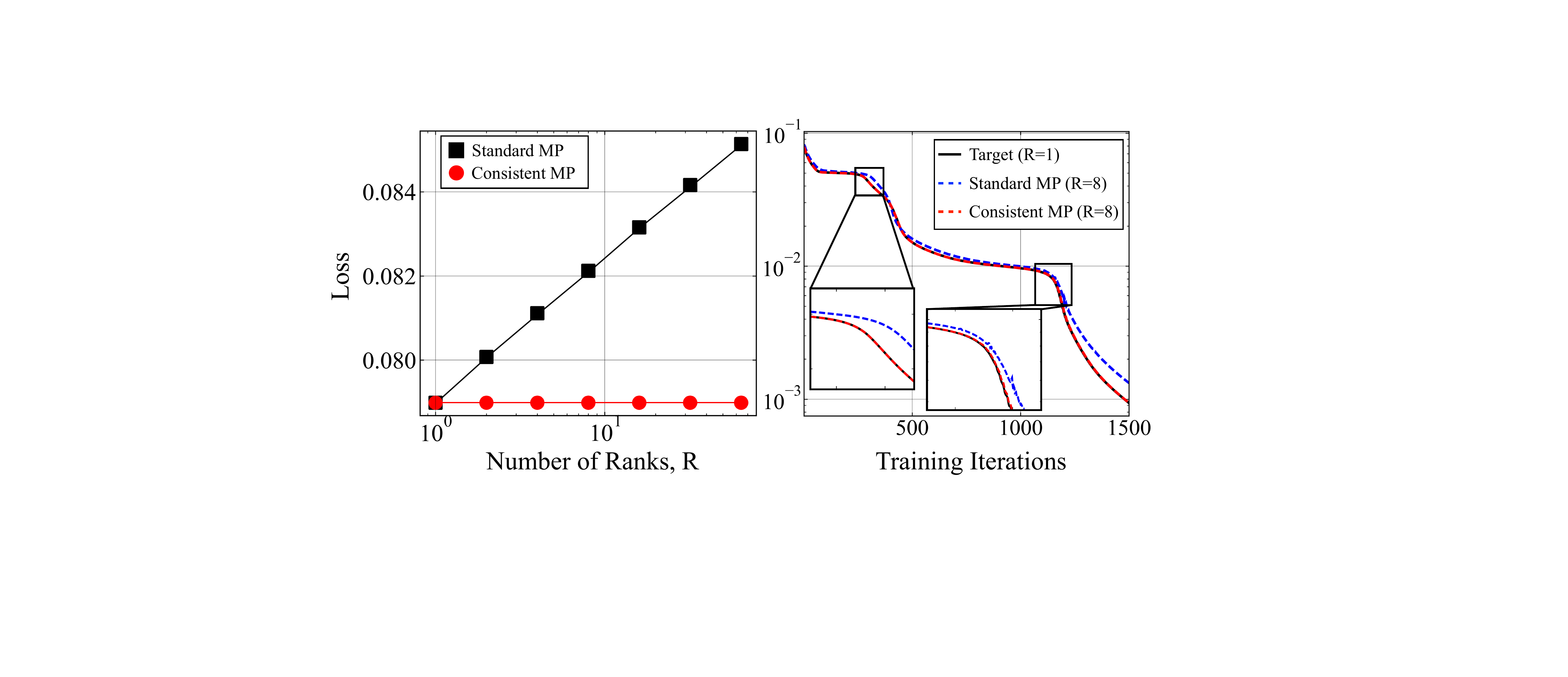}
    \caption{\textbf{(Left)} Loss versus number ranks/sub-graphs $R$ (up to $R=64$) using outputs produced by a randomly-initialized GNN. Black squares indicate usage of standard NMP layers (no halo exchanges), red circles indicate consistent NMP layers (with halo exchanges). \textbf{(Right)} Loss versus training iterations for the target R=1 unpartitioned GNN (solid black), a distributed GNN using standard NMP layers (R=8, dashed blue), and a distributed GNN using consistent NMP layers (R=8, dashed red). For both plots, mesh-based graphs coincide with a cubic spatial domain discretized by $32^3$ elements at the $p=1$ level (see Figs.~\ref{fig:element_graphgen}, \ref{fig:dist_graphgen}).}
    \label{fig:consistency_verif}
\end{figure}

Figure~\ref{fig:consistency_verif} provides a verification of the consistent GNN architecture with respect to Eqs.~\ref{eq:consistency_1} and \ref{eq:consistency_2}. The figure compares distributed GNN outputs in the ``small" configuration from Table~\ref{table:models} using standard NMP layers (no halo exchanges) and consistent NMP layers (with halo exchanges using any of the A2A, N-A2A, or Send-Recv implementations described above).  

Specifically, Fig.~\ref{fig:consistency_verif} (left) shows the impact of the consistent formulation in terms of sensitivity of loss function evaluations to the number of ranks $R$ (or equivalently GPUs) used to decompose the unpartitioned $R=1$ mesh-based graph. At all values of $R$, the MSE loss described in Sec.~\ref{sec:consistent_loss} is evaluated using random model parameters, and by taking the target value as the input for demonstration purposes (the loss is evaluated by taking $\widehat{\bf Y}_r = {\bf X}_r$ in Eq.~\ref{eq:loss_consistent}). The figure, which confirms Eq.~\ref{eq:consistency_1} is being satisfied in a model inference setting, illuminates two key trends: (1) through halo exchanges, the consistent NMP layers remove dependence of GNN outputs on the partitioning procedure, allowing model computations at $R>1$ to be arithmetically equivalent to those at $R=1$; and (2) the non-ideal deviation of the standard NMP formulation (no halo exchanges) from the target $R=1$ solution increases with $R$ at a roughly linear rate. The latter trend is a reflection of the fact that a greater number ranks -- or a greater number of sub-graph partitions -- results in a higher proportion of boundary nodes relative to non-boundary nodes, which in turn creates higher errors in GNN evaluations in the absence of halo exchanges. 

On the other hand, Fig.~\ref{fig:consistency_verif} (right) showcases consistency \textit{during the training phase}. The modeling application is the same as in the left plot: for demonstrative purposes, the target is itself the input, equivalent to an autoencoding task at the node level (the trends, however, are a core property of the formulation and are independent of the modeling application, which is arbitrary for consistency demonstration). Since each iteration involves a forward pass, backward pass, and parameter update, full consistency requires proper satisfaction of Eq.~\ref{eq:consistency_2}, which amounts to equality in the backpropagated gradients of GNN parameters between $R>1$ and $R=1$. Figure~\ref{fig:consistency_verif} shows how the differentiable halo exchange implementations allow for the consistent message passing-based GNN to recover the same optimization curves as the baseline $R=1$ unpartitioned graph in settings where $R>1$, with deviations seen in the inconsistent case for which standard NMP layers without halo exchange synchronization is utilized. 

\subsection{Scaling Analysis}
\label{sec:scaling} 
The previous section demonstrated and verified the physical consistency properties provided by the methodology described in Sec.~\ref{sec:methods}. To build on this, the goal of this section is to address the following questions: what is the cost of enforcing consistency in terms of the halo exchange communication penalties, and what are the model scalability limits? 

To this end, a series of performance and scalability tests were performed on the Frontier supercomputer at the Oak Ridge Leadership Computing Facility \cite{frontier}. 
Frontier is an HPE Cray EX system with 9,408 compute nodes. Each node consists of one 64-core AMD 3rd Gen EPYC CPU and four AMD MI250X GPU, each with 2 Graphics Compute Dies (GCDs) with 64 GB of high-bandwidth memory. Each GCD can be considered a separate GPU for a total of 8 GCDs per node, allowing applications to run with 8 MPI processes per node. CPU-GPU and GPU-GPU connections are achieved with Infinity Fabric, and each node is connected to the Slingshot 11 Interconnect with four HPE Slingshot 25 GB/s network interface cards (NICs).
To offload the computations to the AMD GPU, PyTorch version 2.2.2 was used on Frontier with support for ROCm 5.7.1 and the RCCL collective library version 2.17.1. For GNN-specific functionality, we leveraged PyTorch Geometric 2.5.3. 

The model configurations used in the scaling experiments are summarized in Table~\ref{table:models}, which defines small and large versions of the GNN created by changing the number of parameters in each MLP. Since the number of parameters generally correlates with the model's expressiveness, these versions can be thought of as being tuned for simpler and more complex system dynamics. Moreover, the hidden channel dimensionality directly affects the buffer sizes of the halo exchanges, therefore impacting the cost of enforcing consistency.

Each model configuration was run with two loadings, expressed in terms of the number of local (non-halo) nodes in each sub-graph. This quantity, equivalent to the number of local nodes per MPI rank, which was kept approximately constant at 256k and 512k. The experiments presented here are thus weak scaling tests for the mesh-based consistent GNN model. The sub-graphs were generated by partitioning a NekRS cubic mesh with hexahedral elements of $p=5$ polynomial order following the methods outlined in Sec.~\ref{sec:methods}. Similarly to Sec.~\ref{sec:consistency_demo}, we set $\widehat{\bf Y}_r = {\bf X}_r$, where ${\bf X}_r$ is the velocity vector at each node for some time $t$ of the Taylor Green Vortex solution computed by NekRS. 

\begin{table}[!ht]
  \centering
  \caption{Statistics of Partitioned Sub-Graphs with nominally 512k local nodes of loading (statistics are per-rank)}
  \begin{tabular}{|p{0.7cm}||p{2.3cm}|p{2.2cm}|p{1.95cm}|}
    \hline
    \textbf{Ranks}& \textbf{Graph Nodes} ($10^3$)& \textbf{Halo Nodes} ($10^3$)& \textbf{Neighbors} \\ 
    \textbf{}& \textbf{(min, max, avg)} & \textbf{(min, max, avg)} & \textbf{(min, max, avg)} \\ \hline
    \hline
    8 & 518, 518, 518 & 12.8, 12.8, 12.8 & 2, 2, 2\\ \hline
    64 & 540, 540, 540 & 57.6, 57.6, 57.6 & 11, 11, 11\\ \hline
    512 & 528, 544, 533 & 32.6, 67.6, 44.7 & 5, 15, 7\\ \hline
    2048 & 540, 540, 540 & 57.6, 57.6, 57.6 & 11, 11, 11\\ \hline
  \end{tabular}
  \label{table:gnn_stats}
\end{table}

More details of the partitioned sub-graphs and their variation with the number of ranks are included in Table~\ref{table:gnn_stats} for the case with nominally 512k nodes-per-partition. 
It can be seen that by exploiting the NekRS domain partitioner, the sub-graphs are well balanced for the entire range of MPI ranks tested, with the number of halo nodes and number of neighbors being bounded (the initial increase in the number of halo nodes and neighbors above 8 ranks is due to the domain decomposition strategy changing from vertical rectangular chunks of the domain to sub-cubes).

\begin{figure}
    \centering
    \includegraphics[width=\linewidth]{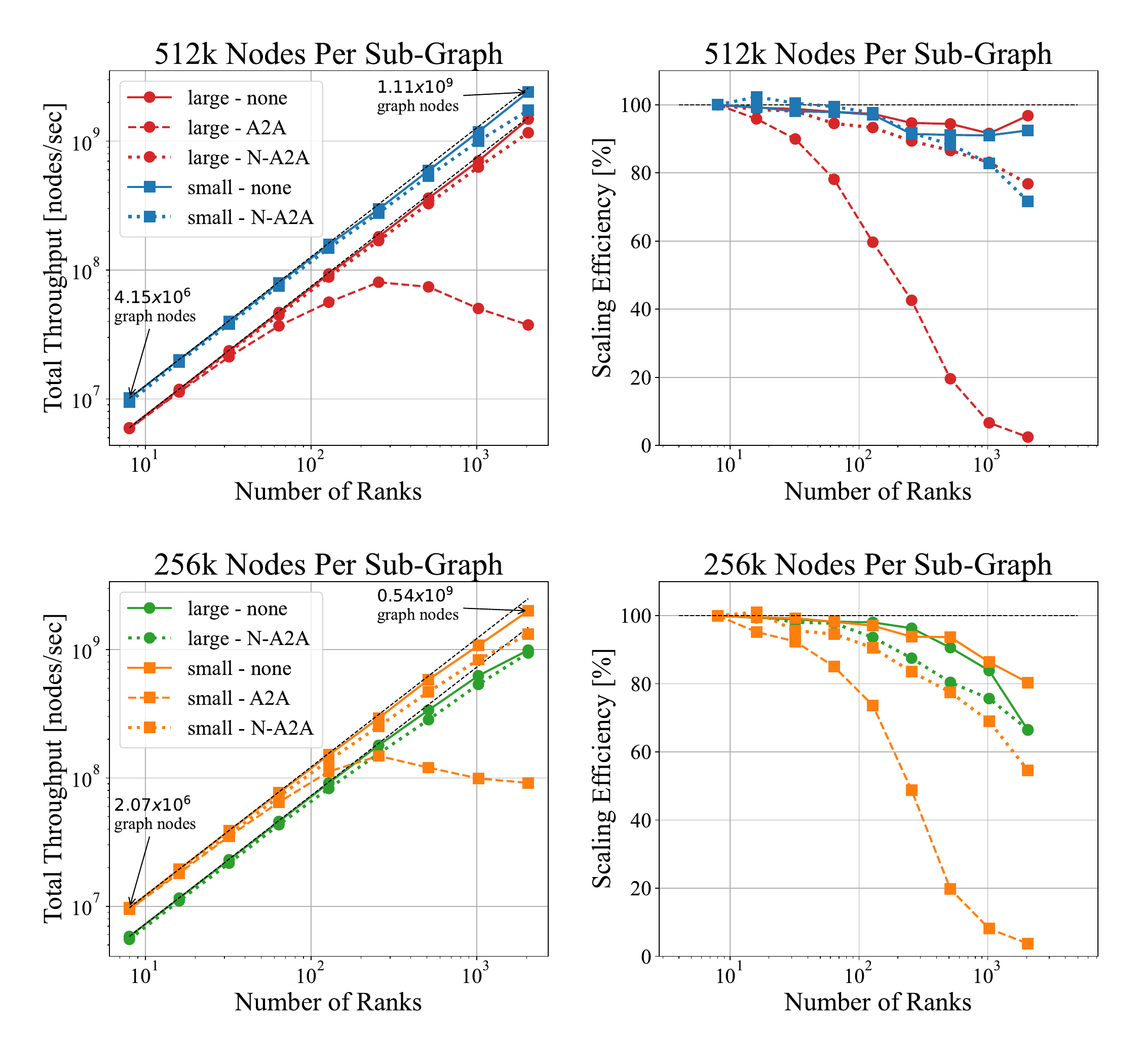}
    \caption{Distributed training weak scaling experiments of the mesh-based consistent GNN on the Frontier supercomputer with the configurations described in Table~\ref{table:models}. Time used in total throughput corresponds to one full training iteration. The size of the total graph grows linearly with the number of ranks, from $4.15 \times 10^6$ nodes at 8 ranks ($R=8$) to $1.105\times10^9$ at 2048 ranks ($R=2048$).}
    \label{fig:scaling}
\end{figure}

\textbf{It should be noted that since the loading is almost constant on a per-rank (or per sub-graph) basis, the size of the total graph grows linearly with the number of ranks, from $4.15 \times 10^6$ nodes at 8 ranks to $1.105\times10^9$ at 2048 ranks.} The ability to efficiently operate on graph sizes on the order of $10^9$ nodes (see Fig.~\ref{fig:scaling}), while maintaining arithmetic consistency in evaluations, highlights the strength of the method presented in this work. As mentioned in Sec.~\ref{sec:intro}, scaling up to the billion-node mark (and beyond) for this application is crucial for developing models using modern high-fidelity unstructured simulation tools.

Figure~\ref{fig:scaling} shows the total training throughput, measured as the total number of graph nodes processed per second in one training iteration across all ranks. Also shown in Fig.~\ref{fig:scaling} is the weak scaling efficiency for the various configurations and loadings of the mesh-based consistent GNN model. Specifically, the plots show scale-out performance from 8 MPI ranks (1 Frontier node) to 2048 MPI ranks (256 Frontier nodes). 

Starting with the inconsistent model which does not perform the halo exchange, both model sizes obtain weak scaling efficiency above 90\% up to the 2048 MPI ranks tested for the larger loading case. It is worth mentioning that enabling the use of all four NICs on the Frontier nodes for PyTorch DDP (with \verb|NCCL_SOCKET_IFNAME=hsn0,hsn1,hsn2,hsn3|) and enabling RCCL to use libfabric (through the AWS-OFI-RCCL plugin) were key in obtaining such high efficiency. As expected, however, the scaling efficiency deteriorates for the smaller loading case, particularly beyond 512 MPI ranks (64 Frontier nodes) where the communication cost of the AllReduce calls for the consistent loss and backward pass become more prominent. 

For the consistent model formulation which does perform the halo exchange, Fig.~\ref{fig:scaling} usefully illustrates the inefficiencies of the standard A2A implementation. The drop in scaling efficiency is particularly prominent for the large model configuration and larger loading due to the buffer sizes depending directly on the number of hidden channels and number of halo nodes. As noted earlier in this section, the standard A2A is a naive implementation of the halo exchange, since each rank only needs to communicate with a few neighbors (see Table~\ref{table:gnn_stats}); the poor scalability is therefore expected. By contrast, the Neighbor-A2A implementation of the halo exchange shows adequate scaling with the larger loading case, significantly outperforming the standard A2A implementation and demonstrating the viability of the consistent formulation proposed in this work. Once again, we observe that reducing the size of the sub-graphs has a negative impact on the scaling efficiency. 
Focusing on the effects of model size, only slight differences in the scaling efficiency are observed for the larger loading case, whereas the smaller model shows a noticeably reduced performance beyond 512 MPI ranks.

\begin{figure}
    \centering
    \includegraphics[width=\linewidth]{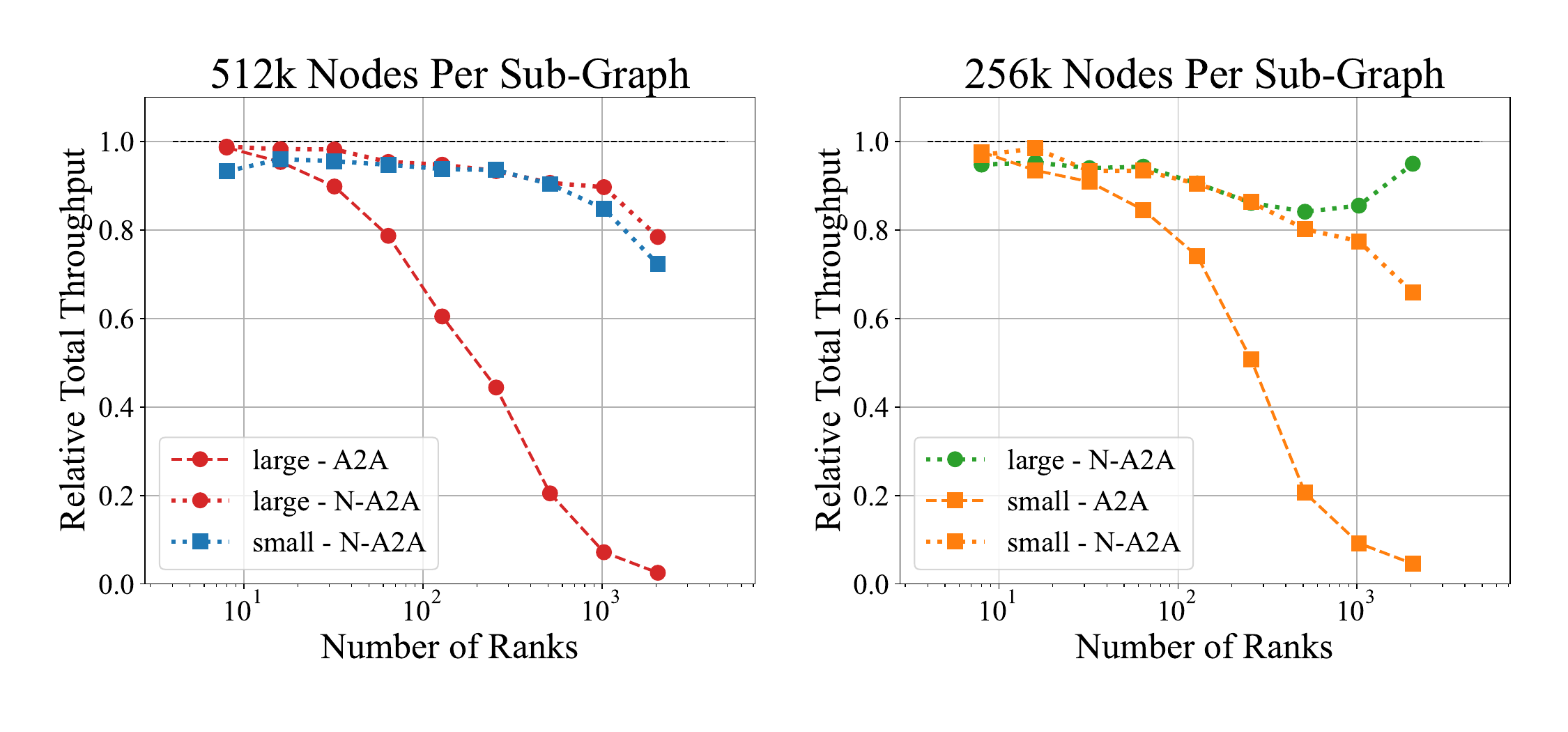}
    \caption{Distributed training throughput of the mesh-based consistent GNN with the halo exchange relative to the inconsistent model without halo exchange (up to $R=2048$ on the x-axis), isolating the additional cost of the AllToAll implementations.}
    \label{fig:rel_scaling}
\end{figure}

The performance impacts of ensuring algebraic consistency with the halo exchange in the forward and backward passes are highlighted further in Fig.~\ref{fig:rel_scaling}, where the total parallel training throughput obtained with the A2A and N-A2A implementations is normalized by the one obtained with no halo exchange. Effectively, this normalization identifies the cost of the 8 \verb|all_to_all| communications performed each training step (one \verb|all_to_all| needs to be performed for each \textcolor{black}{neural} message passing layer in the forward and backward passes). 
While the added cost of enforcing consistency with the standard A2A implementation quickly becomes impractical as the number of MPI ranks increases, the N-A2A implementation only adds a marginal cost to the model training. In fact, the relative throughput for both model sizes on the larger sub-graphs remains above 0.95 until 64 ranks, and with the larger model size above 0.9 until 1024 ranks (i.e., a drop in training throughput of 10\%). More investigation is needed to better understand the increase in cost of the halo exchange above 20\% at 2048 ranks. 

Consistent with Figure~\ref{fig:scaling}, with the smaller sub-graphs the \verb|all_to_all| communication cost negatively impacts scaling efficiency more noticeably and the relative throughput drops below 0.9 beyond 128 MPI ranks (16 Frontier nodes). This behavior is especially noticeable with the small model size at larger scale, even though this configuration produces the smallest send and receive buffers for the \verb|all_to_all|.

\section{Conclusion}
This work develops a distributed graph neural network (GNN) approach for mesh-based modeling applications in the \textcolor{black}{neural} message passing (NMP) framework. The methodology relies on two core components: (1) interoperability with existing high-fidelity physics simulation tools, and (2) enforcement of physical consistency in \textcolor{black}{neural} message passing operations on distributed graphs. Regarding the first component, the mesh-based graph (alongside the mesh partitioner) is derived from discretizations used in the NekRS CFD solver. Regarding the second component, a consistent NMP layer and loss are developed. Using halo exchanges over graph nodes, the consistent NMP layers ensure that (a) GNN operations on multi-GPU distributed graphs are arithmetically equivalent to those on unpartitioned graphs, and (b) evaluations are invariant to both the number and location of sub-graph boundaries. 
Consistency of the mesh-based GNN was demonstrated using 8 MPI ranks (i.e., 8 sub-graphs) for 1,500 training iterations.

The performance of the method under different loadings (i.e., number mesh nodes per sub-graph) and model sizes was presented through a series of weak scaling experiments reaching 2048 MPI ranks ($\mathcal{O}(10^9)$ graph nodes) on the Frontier supercomputer. These experiments also compared two consistency approaches: (1) standard AllToAll (A2A), and (2) Neighbor-AllToAll (N-A2A) where the send and receive buffers are strategically created to effectively reduce the \verb|torch.distributed.nn.all_to_all| operation to a neighbor SendReceive communication. 
While the added cost of enforcing consistency with the standard A2A implementation becomes impractical at large scale, the N-A2A implementation was shown to only add a marginal cost to model training, demonstrating the efficiency of the consistent formulation proposed in this work for developing GNN models using modern high-fidelity unstructured simulation data.

The scope of this study focuses on introducing the method itself and analyzing its scaling properties on large graphs. As such, there are multiple avenues for future work. A direct extension is to apply the consistent GNNs developed here to create more realistic surrogate and closure models of multi-physics fluid flows using data stored on very large meshes. Another direction is to leverage scalable workflow tools for in-situ training, which casts the high-fidelity physics simulation (like NekRS) as a data generator without ever writing to disk, thereby enhancing modeling capabilities \cite{riccardo_smartsim,Maric_2024}. On the scaling front, additional directions include executing similar tests shown in this work on different supercomputers. Since the consistent GNN scalability is a function of halo exchange buffer sizes, NMP arithmetic intensity, and graph size, it offers a unique and complex benchmark for comparing performance across many HPC platforms. The authors are pursuing these directions and hope to report on them in future studies.

\section{Acknowledgments}
{\color{black} The manuscript has been created by UChicago Argonne, LLC, Operator of Argonne National Laboratory (Argonne). The U.S. Government retains for itself, and others acting on its behalf, a paid-up nonexclusive, irrevocable world-wide license in said article to reproduce, prepare derivative works, distribute copies to the public, and perform publicly and display publicly, by or on behalf of the Government. This work was supported by the U.S. Department of Energy (DOE), Office of Science under contract DE-AC02-06CH11357. SB and PP acknowledge laboratory-directed research and development (LDRD) funding support from Argonne's Advanced Energy Technologies (AET) directorate through the Advanced Energy Technology and Security (AETS) Fellowship. RM acknowledges funding support from DOE Advanced Scientific Computing Research (ASCR) program through DOE-FOA-2493 project titled “Data-intensive scientific machine learning”. RBK acknowledges support by the Office of Science, U.S. Department of Energy, under contract DE-AC02-06CH11357. This research used resources of the Argonne Leadership Computing Facility (ALCF) at Argonne National Laboratory, which is a U.S. Department of Energy Office of Science User Facility operated under Contract No. DE-AC02-06CH11357. This research also used resources of the Oak Ridge Leadership Computing Facility (OLCF) at the Oak Ridge National Laboratory, which is supported by the Office of Science of the U.S. Department of Energy under Contract No. DE-AC05-00OR22725.}
\textcolor{black}{Finally, the authors acknowledge Josh Romero and Thorsten Kurth of the NVIDIA Corporation for the insightful discussions about the neighbor all-to-all implementation of the halo exchange.}

\printbibliography

\end{document}